\documentclass{nature}

\spacing{1}
\usepackage{graphicx}
\usepackage{nature_journal_style}

\publishedat{doi:10.1038/nature07266}{http://www.nature.com/nature/journal/v455/n7213/full/nature07266.html}
\title{Clustered star formation as a natural explanation of 
  the H$\alpha$~cutoff in disc galaxies}
\author{Jan Pflamm-Altenburg$^{1}$ \& Pavel Kroupa$^{1}$}
\begin{document}

\maketitle

\begin{affiliations}
\item Argelander-Institut f\"ur Astronomie (AIfA),
  Universit\"at Bonn, 53121 Bonn, Germany

\end{affiliations}
\begin{abstract}
  Star formation is mainly determined by the observation of H$\alpha$
  radiation which is related to the presence of short lived massive
  stars. Disc galaxies show a strong cutoff in H$\alpha$ radiation at a
  certain galactocentric distance which has led to the conclusion that star
  formation is suppressed in the outer regions of disc galaxies. This is 
  seemingly in contradiction to recent 
  UV observations\cite{boissier2007a} that imply disc galaxies to have
  star formation beyond the H$\alpha$ cutoff and that 
  the star-formation-surface  
  density is linearly related to the underlying gas surface density being 
  shallower than derived from H$\alpha$ luminosities\cite{kennicutt1998a}. 
  In a galaxy-wide formulation
  the clustered nature of star formation has recently 
  led to the insight that the total
  galactic H$\alpha$ luminosity is non-linearly related to the galaxy-wide
  star formation rate\cite{pflamm-altenburg2007d}. Here we show that a 
  local formulation of the concept of clustered star formation naturally leads
  to a steeper radial 
  decrease of the H$\alpha$ surface luminosity than the 
  star-formation-rate surface density
  in quantitative agreement with the observations, and that the observed 
  H$\alpha$ cutoff arises naturally. 
\end{abstract}

The integrated galactic initial mass function (IGIMF) describes the 
mass spectrum of all newly formed stars in a galaxy. The IGIMF is
calculated by adding up all stars of all newly formed star 
clusters\cite{vanbeveren1983a,weidner2005a} 
and falls off more steeply with increasing stellar masses for massive
stars\cite{weidner2005a} than the canonical initial 
mass function (IMF) in each star cluster due to the combination
of two effects: the masses of the young star clusters are 
distributed according to the embedded cluster mass function (ECMF), for which
the upper mass limit is a function of the 
total star formation rate\cite{weidner2004b} (SFR), and the stellar 
upper mass limit
of the IMF is a function of the total star cluster mass\cite{weidner2004a}. 
Consequently, the total fraction of massive stars and 
therefore the total H$\alpha$ luminosity drops faster with decreasing SFR
than the SFR\cite{pflamm-altenburg2007d}.
The IGIMF theory has already been shown to naturally lead to the 
observed mass-metallicity relation of galaxies\cite{koeppen2007a} and has
received recent empirical verification in a study of IMF variations among 
galaxies\cite{hoversten2008a}.

In order to construct a quantitative local IGIMF-theory we introduce
the local embedded cluster mass function (LECMF),
\begin{equation}\xi_{\mathrm{LECMF}}(M_\mathrm{ecl},x,y)=\frac{d
    N_\mathrm{ecl}}{d M_\mathrm{ecl}\;d x\;d y}\;,\end{equation} which
defines the number of newly formed star clusters with mass
$M_\mathrm{ecl}$ per unit area at the location $x,y$ in a disc galaxy.
Observations\cite{lada2003a} of Galactic star forming regions show
that this function is a single part power law,
$\xi_\mathrm{LECMF}\propto M_\mathrm{ecl}^{-\beta}$, with an index of
$\beta=2$. The smallest cluster mass\cite{weidner2005a},
$M_\mathrm{ecl,min} = 5\;M_\odot$, should form at any place in the
galaxy, whereas the most massive star cluster,
$M_\mathrm{ecl,max,loc}(x,y)$, which can form locally is expected to
depend on the local gas density, i.e. how much material is locally
available for star cluster formation.
Observations\cite{weidner2004b,weidner2005a} show that the most
massive star cluster, $M_\mathrm{ecl,max}$, of the whole galaxy is a
function of the total galactic star formation rate. To express the
upper limit of the LECMF in dependence on the local gas surface
density we write
\begin{equation}\label{eq_m_ecl_max_loc}M_\mathrm{ecl,max,loc}(x,y)=
M_\mathrm{ecl,max}\left(\frac{\Sigma_\mathrm{gas}(x,y)}
{\Sigma_\mathrm{gas,0}}\right)^\gamma\;,
\end{equation}
where $\Sigma_\mathrm{gas}(x,y)$ and $\Sigma_\mathrm{gas,0}$
are the gas density at the location $x,y$ and at the origin,
respectively.  The local mass of all star clusters between the two
mass limits is determined by the local star formation rate which is
described by the
Kennicutt-Schmidt-law\cite{kennicutt1998a,kennicutt2007a}
$\Sigma_\mathrm{SFR}(x,y)= A\;\Sigma_\mathrm{gas}^{N}(x,y)\;$ with
$N=1.4$ being the widely accepted value\cite{kennicutt1998a}, whereas
$N=0.99$ follows from recent UV observations\cite{boissier2007a}.  As
UV emission is a star formation tracer that is much less sensitive to
the presence of OB stars than H$\alpha$ emission, the true exponent
$N$ must be much closer to the value derived from UV
observations. Thus, we chose $N=1$.  Interestingly, the high
gas-density part of the Kennicutt-Schmidt-plot\cite{kennicutt1998a}
based on FIR observations also has a flatter slope of $N=1.08$ and
galaxy evolution models suggest $N$ not to exceed unity in order to
reproduce observed radial density profiles of disc
galaxies\cite{zasov2006a}, confirming our choice.  The mass spectrum
of all newly formed stars per unit area (the local integrated galactic
initial mass function, LIGIMF) is calculated by adding up all newly
formed stars of all young star clusters, and the H$\alpha$ surface
density follows by adding up the H$\alpha$ flux contributions of all
newly formed stars.  The newly formed stars in each young star cluster
are distributed according to the invariant canonical initial mass
function (IMF)\cite{kroupa2001a,kroupa2002a} with a fixed lower mass
limit but an upper mass limit depending on the total cluster
mass\cite{weidner2004a}. Young star clusters above
$\approx$3000~$M_\odot$ have constant H$\alpha$-light-to-mass ratios
whereas smaller clusters are increasingly H$\alpha$
under-luminous\cite{pflamm-altenburg2007d}.  With decreasing star
formation rate surface density the upper mass limit of the LECMF
decreases, and consequently the fraction of under-luminous star
clusters increases. Thus UV and H$\alpha$ scale differently with the
star formation surface density, gas surface density or galacotentric
radius, respectively. A detailed explanation how the H$\alpha$ surface
luminosity is calculated is given in the Supplementary Discussion.

The LIGIMF-theory is next applied to a sample of disc 
galaxies\cite{kennicutt1989a} with meassured gas surface
densities and H$\alpha$ surface luminosities of H~{\sc ii} regions
averaged over annuli at different galactocentric radii. It is known that 
ionising photons emitted by massive stars can escape from well defined
H~{\sc ii} regions and lead to recombinations and thus H$\alpha$ radiation
in the surrounding diffuse ionised gas (DIG)\cite{ferguson1996a}.
Using  H$\alpha$~emission as a star formation tracer this kind of  
photon leakage  has to be taken into account to get an estimate
of the true star formation rate. The study\cite{ferguson1996a} 
of NGC~247 and NGC~7793 allows to construct a correction procedure
in order to obtain the total H$\alpha$ surface luminosities from the 
surface luminosities of H~{\sc ii} regions only (see Supplementary
Discussion).  

\begin{figure}
  \includegraphics{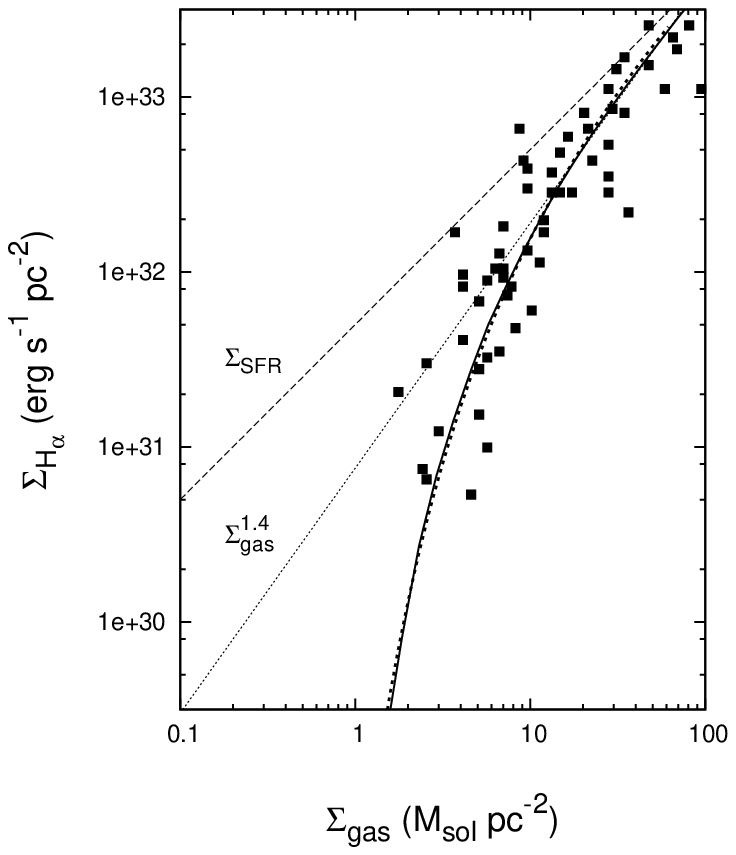}
  \caption{H$\alpha$-luminosity
    surface density versus total gas surface density.}{\label{kennicutt_schmidt_fig}
    The H$\alpha$-luminosity
    surface density versus the total gas surface density observed 
    for seven disc galaxies\cite{kennicutt1989a} averaged over 
    annuli at different galactocentric radii is plotted (black squares) 
    after correcting  for photon leakage
    from H~{\sc ii} regions (see Supplementary Discussion).
    These galaxies  have a mean star formation rate of 
    SFR=6.9~$M_\odot$~yr$^{-1}$ 
    (3.2 -- 16.4~$M_\odot$~yr$^{-1}$)\cite{kennicutt1989a,kennicutt1998a}, 
    a mean total gas mass of $M_{\mathrm{gas}}=2.1\cdot 10^{10}\;M_\odot$ 
    (0.6 -- $3.6 \cdot 10^{10}\;M_\odot$)\cite{kennicutt1989a,kennicutt1998a} 
    and a mean scale length of $r_\mathrm{d}=4.4$~kpc 
    (3.9 -- 5.2~kpc)\cite{kenney1989a,wong2002a,schuster2007a,crosthwaite2007a}.
    These mean values define our model standard disc galaxy.
    For a choice of $\gamma=\frac{3}{2}$ the LIGIMF-theory predicts an
    $\Sigma_\mathrm{H\alpha}$-$\Sigma_\mathrm{gas}$ relation which matches
    the observations excellently (solid line). Note that
    the underlying true star-formation density as derived from UV
    observations\cite{boissier2007a}
    is directly proportional to the gas surface density ($N=1$) and 
    is shown after converting it into an H$\alpha$ surface 
    luminosity using the wrong linear Kennicutt H$\alpha$-SFR 
    relation\cite{kennicutt1998a,kennicutt1994a} (dashed line) and
    shows the expected $\Sigma_\mathrm{H\alpha}$-$\Sigma_\mathrm{gas}$
    relation based on the classical picture which is in disagreement with 
    the observations. Furthermore, the 
    H$\alpha$-luminosity surface density in the high-luminosity part 
    ($\Sigma_{\mathrm{H}\alpha}\ge$ 10$^{32.5}$ erg s$^{-1}$ pc$^{-2}$)
    depends, for the correct LIGIMF-theory,
    on the gas surface density with a power of 1.4 (dotted
    line, extrapolated to low H$\alpha$ surface luminosity) in agreement
    with the classical Kennicutt-Schmidt slope of $N=1.4$.
    The LIGIMF-theory puts the hitherto inconsistent H$\alpha$ and UV 
    observations in perfect agreement with each other. Both the steeper 
    high-luminosity slope of $N=1.4$ and the H$\alpha$ cutoff at low gas
    densities are  two simultaneous outcomes of the LIGIMF-theory.
    The thick dotted line which coincides with the thick solid curve shows
    a fitting function of the LIGIMF-model (see Supplementary Discussion).}
\end{figure}

\begin{figure}
  \includegraphics{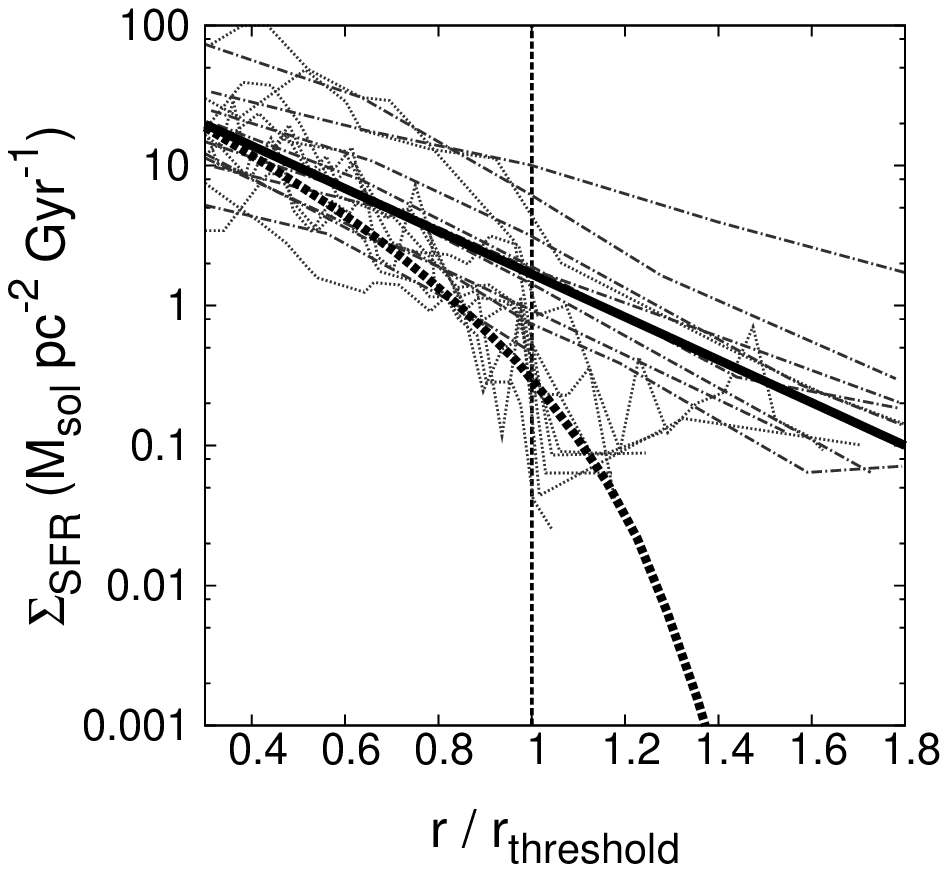}
  \caption{Star formation surface density versus scaled galactocentric radius.}{\label{kennicutt_radial_fig}Radial distribution of the star
    formation surface density of nine disc galaxies 
    based on UV\cite{boissier2007a} 
    (thin dashed-dotted lines) 
    and H$\alpha$\cite{martin2001a} (thin dotted lines) 
    observations that rely on a wrong linear 
    conversion\cite{kennicutt1994a} 
    between the corresponding H$\alpha$-luminosity surface
    density and star-formation-rate surface density after correction for
    photon leakage (see Supplementary Discussion).  
    The galactocentric radius is in units of the 
    H$\alpha$ threshold radius\cite{martin2001a}. Over-plotted is
    the true underlying star-formation-rate surface density of our standard
    disc galaxy (thick black line) 
    as defined in Fig.~\ref{kennicutt_schmidt_fig} 
    and the model H$\alpha$ surface 
    luminosity (thick dotted black line) converted into a star-formation 
    surface density using 
    the same linear conversion\cite{martin2001a,kennicutt1994a}. The LIGIMF-theory thus
    naturally accounts for the discrepant $\Sigma_\mathrm{SFR}$ value at
    a particular radius.}
\end{figure}

For a linear star-formation law ($N=1$) as derived from UV
observations\cite{boissier2007a} the LIGIMF-theory
predicts an H$\alpha$ surface luminosity as a function of the gas surface
density which is in full agreement with the 
observations (Fig.~\ref{kennicutt_schmidt_fig}). Additionally, the
radial H$\alpha$ profile derived in the LIGIMF-theory matches
the observations perfectly (Fig.~\ref{kennicutt_radial_fig}). 
The concept of clustered
star formation resolves the discrepancies between H$\alpha$ and UV
observations completely.

At first sight it might be objected that the LIGIMF-theory contradicts
observations of the UV sources in the outer disc of galaxies:
5\%--10\% of all clusters in the outer discs of galaxies detected in
UV have associated H$\alpha$ emission\cite{zaritsky2007a}.  The age
estimates of the UV knots range up to 400~Myr. Clusters with H$\alpha$
emission have ages $\le$20~Myr as they are powered by short lived
massive stars and therefore 5\% of all observed UV knots are expected
to have associated H$\alpha$ emission in agreement with
observations. The LIGIMF-theory predicts an overabundance of
under-luminous star clusters beyond the H$\alpha$ cutoff and a smaller
number ratio of H$\alpha$ to non-H$\alpha$-emitting UV knots is
expected.  But under-luminous does not mean no H$\alpha$ emission. In
the LIGIMF theory each young UV cluster is an H$\alpha$ source, too,
but UV and H$\alpha$ luminosity scale differently with the cluster
mass.  Thus, this finding\cite{zaritsky2007a} is entirely consistent
with the LIGIMF-theory.  The observed UV knots in the outer disc of
M83 are at any time systematically smaller than their counterparts in
the inner disc\cite{thilker2005a} in agreement with the fundamental
basics of the LIGIMF-theory. Even one outstanding massive young star
cluster in the outer region of M83 does not contradict this principle
but is expected from a statistical point of view (see Supplementary
Discussion). Furthermore, the M83 FUV luminosity function of
outer-disc stellar complexes is steeper than for the inner-disc
population\cite{thilker2005a}.  A similar trend is reported in NGC~628
for the H$\alpha$ luminosity function of H~{\sc ii}
regions\cite{lelievre2000a}.  In the LIGIMF-theory inner disc LECMFs
have higher upper mass limits than in the outer disc
LECMFs. Integration of the LECMFs over the outer regions leads to a
steeper ECMF of the outer disc than the resultant ECMF of the inner
disc, indicating that outer disc star formation complexes are
systematically smaller than the inner disc ones. This integration
effect is fundamentally the same as the IGIMF being steeper for dwarf
galaxies with low global star formation rates than for disc galaxies
with high star formation rates\cite{pflamm-altenburg2007d}.

Previously, the H$\alpha$ cutoff has been
explained\cite{kennicutt1989a} by a drop of the local gas density
below a critical density determined by the stability condition of a
thin isothermal disc\cite{toomre1964a,cowie1981a} where no star
formation can occur. In contradiction to this explanation recent UV
observations\cite{boissier2007a} reveal star formation outside the
H$\alpha$ cutoff, and dwarf galaxies\cite{hunter2001a} show star
formation although their average gas density is lower than the
critical density. Indeed, it has been shown that in regions with
densities lower than the critical value, star formation can be driven
by other than thermal instabilities\cite{elmegreen2006a}. It has been
argued that H~{\sc ii} regions powered by the same massive stars are
larger in a thin environment, i.e. at large galactocentric radii, than
in a dense one and thus identical H~{\sc ii} regions become fainter in
the outer galaxy. Therefore, it has been
concluded\cite{elmegreen2006a} that the H$\alpha$ surface luminosity
should drop faster than the star formation rate surface
density. Indeed, the surface brightness of individual H~{\sc ii}
regions should be fainter in the outer galaxy.  But the H$\alpha$
surface density considered in star formation laws refers to the total
H$\alpha$ luminosity per unit area of the galaxy and not to the cross
section of the H~{\sc ii} region. Identically powered H~{\sc ii}
regions contribute equally to the H$\alpha$ surface luminosity
independently of their location in a thin or a dense gas
environment. Thus, the proposed solution\cite{elmegreen2006a} does
neither explain the H$\alpha$ cutoff nor the different slopes of the
UV-based and H$\alpha$-based star formation laws.  It has been shown
recently that a required minimum column density for massive star
formation might exist\cite{krumholz2008a} implying star formation with
no massive stars in low density environments. However, this model
predicts a top-heavy IMF for cloud column densities much larger than
this threshold, for which no observational evidence
exists\cite{kroupa2002a}, and allows no quantitative linkage of
H$\alpha$ luminosity and the star formation rate.  Contrary to this
previously existing work, the LIGIMF theory is in excellent agreement
with the observed radial H$\alpha$- and UV-luminosity profiles
(Fig.~\ref{kennicutt_radial_fig}) and the Kennicutt-Schmidt star
formation law (Fig.~\ref{kennicutt_schmidt_fig}), and also allows the
determination of star formation rates even in H$\alpha$-faint galaxy
regions.

\bibliographystyle{naturemag}
\bibliography{imf,star-formation,disk_galaxies,astro-ph,cmf,dwarf_galaxies}

\noindent{\bf \sffamily Supplementary Information} is linked to the
online version of the paper at www.nature.com/nature.

\noindent{\bf \sffamily Acknowledgements} We thank Klaas S. de Boer for 
stimulating discussions.

\noindent{\bf \sffamily Author Information} Reprints and permissions
information is available at www.nature.com/reprints.
Correspondence and request for
materials should be addressed to J.P.-A. (jpflamm@astro.uni-bonn.de).

\clearpage

\clearpage

\clearpage

{\Large\bfseries\noindent\sloppy \textsf{Supplementary Discussion}}

This Supplementary Discussion provides more information on i) how the
H$\alpha$ surface luminosity is calculated, ii) the correction for
photon leakage from H~{\sc ii} regions, iii) the local applicability
of the IGIMF theory, iv) a discussion on the sizes of typical star
forming regions, v) a discussion of the presence of star clusters more
massive than $M_\mathrm{ecl,max,loc}$ and vi) the fitting formula for
the Kennicutt-Schmidt law in the LIGIMF-model.

{\large\bfseries\noindent\sloppy 
  \textsf{i) Calculation of the H$\alpha$ surface luminosity}}

In order to construct a local IGIMF theory we start with 
the IGIMF theory\cite{weidner2005a,pflamm-altenburg2007d}, 
developed for the entire galaxy, and transfer all quantities 
into their corresponding local surface densities.

Thus, we first define the local embedded cluster mass function (LECMF), 
\begin{equation}
\xi_{\mathrm{LECMF}}(M_\mathrm{ecl},x,y)=
\frac{d N_\mathrm{ecl}}{d M_\mathrm{ecl}\;d x\;d y}\;,
\end{equation}
which is the number of newly formed star clusters with mass
$M_\mathrm{ecl}$ per unit area at the location $x,y$ in a disc galaxy.
Observations\cite{lada2003a} of Galactic star forming regions show
that the LECMF is a single part power law, $\xi_\mathrm{LECMF}\propto
M_\mathrm{ecl}^{-\beta}$, with an index of $\beta=2$. The local mass
surface density of newly formed stars is
\begin{equation}
\delta t\;\Sigma_\mathrm{SFR}(x,y)=
\int_{M_\mathrm{ecl,min}}^{M_\mathrm{ecl,max,loc}(x,y)}\xi_\mathrm{LECMF}{(M_\mathrm{ecl})}\;M_\mathrm{ecl}\;\mathrm{d}M_\mathrm{ecl}\;,
\end{equation}
where $\Sigma_\mathrm{SFR}$ is the star-formation surface density, 
$\delta t\approx10\;\mathrm{Myr}$ is the time span required to
populate the cluster mass function completely\cite{weidner2004b}, and 
$M_\mathrm{ecl,min}=5$~M$_\odot$ is the 
smallest cluster mass\cite{weidner2005a}. 

The observed most massive embedded star cluster in a galaxy is determined 
through the total star formation rate\cite{weidner2004b,weidner2005a},
\begin{equation}
\frac{M_\mathrm{ecl,max}}{M_\odot}=
84793\;\left(\frac{\mathrm{SFR}}{M_\odot\;\mathrm{yr}^{-1}}\right)^{0.75}\;.
\end{equation}
To express the upper mass limit of the LECMF in dependence on the local
gas surface density we write 
\begin{equation}
M_\mathrm{ecl,max,loc}(x,y)=
M_\mathrm{ecl,max}\left(\frac{\Sigma_\mathrm{gas}(x,y)}
{\Sigma_\mathrm{gas,0}}\right)^\gamma\;.
\end{equation}
The maximal gas-mass surface density,
associated with the
position of the most massive embedded star cluster,
 in a disc galaxy,
which has a  total gas mass $M_\mathrm{gas}$ and 
a single-exponential gas disc,
\begin{equation}
\Sigma_\mathrm{gas}(x,y)=
\Sigma_\mathrm{gas,0}\;e^{-r/r_\mathrm{d}}\;,
\end{equation} 
with a scale length $r_\mathrm{d}$, is determined by
\begin{equation}
\Sigma_\mathrm{gas,0}=
\frac{M_\mathrm{gas}}{2\;\pi\;r_\mathrm{d}^2}\;.
\end{equation}

One may raise the objection if the crude assumption of an
exponential gas disc model is sufficient to construct a standard
galaxy as the outer galaxy discs may also be described by other
expressions as for example a power law.  The key issue of the theory
developed here is that the gas surface density generally tends to
decrease with increasing galactocentric radius.  The main aim here is
not the detailed modelling of individual galaxies but to explain the
characteristic discrepancies between H$\alpha$ and UV observation and
their traditional interpretations: i) a distinct star formation cutoff
inferred from the H$\alpha$ cutoff contrary to the extended radial
star formation inferred from UV. ii) A steeper star-formation-law
slope inferred from H$\alpha$ (N=1.4) observations than from UV
(N=1.0) observations within the H$\alpha$ cutoff.  The explanation
through the present LIGIMF theory relies on the description of
the gas surface density beyond the H$\alpha$ cutoff. If the very outer
gas disc is better described by for example a power law rather than by
a single exponential law then this would not change the outcome of
the LIGIMF theory.  The radial H$\alpha$ profiles of the disc
galaxies examined in UV\cite{boissier2007a} and Fig.~2 in the main
Text are from ref.~30.  Figure~9 of ref.~30 shows the total gas
surface densities of some of these galaxies. As this is a log-lin
diagram it can easily be seen that the general trend of how the gas
surface density scales with radius can be well described by an
exponential law up to $\approx$2~H$\alpha$ radii. If the aim is to
construct detailed mass models of individual galaxies then an
exponential law is inaccurate, but by far sufficient for our purpose
here.

The star-formation surface density is described by the 
Kennicutt-Schmidt-law\cite{kennicutt1998a,kennicutt2007a},
\begin{equation}
\label{supp_eq_sfr_law}
\Sigma_\mathrm{SFR}(x,y)=
  A\;\Sigma_\mathrm{gas}^{N}(x,y)\;,
\end{equation}
with $N=1.4$  being the widely accepted  value\cite{kennicutt1998a}, 
whereas $N=0.99$ follows from UV observations\cite{boissier2007a}. 
As UV emission is a star formation tracer that is much
less sensitive to the presence of OB stars than H$\alpha$ emission,
the true 
exponent $N$ must be much closer to the value derived from UV
observations. Thus, we chose $N=1$. 
Interestingly, the high gas-density part of the 
Kennicutt-Schmidt-plot\cite{kennicutt1998a} based on FIR observations 
also has a flatter slope of $N=1.08$, and galaxy evolution models 
suggest $N$ not to exceed unity in order
to reproduce observed radial density profiles of 
disc galaxies\cite{zasov2006a}, confirming our choice. 

The factor $A$ is
determined by integrating Supplementary equation~(\ref{supp_eq_sfr_law})
over the whole disc,
\begin{equation}
A=\frac{\mathrm{SFR}\;n^2}{2\;\pi\;
\Sigma_{\mathrm{gas},0}\;r_\mathrm{d}^2}\;.
\end{equation}

In each embedded star cluster the newly formed stars are distributed
according to the invariant canonical
initial mass function (IMF)\cite{kroupa2001a,kroupa2002a} 
which is a two-part
power law, $m^{-\alpha_\mathrm{i}}$,
with indices $\alpha_1=1.3$ between $m_\mathrm{low}=0.1$ and
$m_1=0.5$~$M_\odot$ , and 
$\alpha_2=2.35$ between $m_1=0.5$~$M_\odot$ and $m_\mathrm{max}$, where
$m_\mathrm{max}=m_\mathrm{max}(M_\mathrm{ecl})$ 
is given by the maximum-stellar-mass--star-cluster-mass 
relation\cite{weidner2004a}. This relation is well defined empirically and
is a result of feedback-driven star formation on a star-cluster spatial 
scale of $\le$ few pc (Weidner, Kroupa \& Goodwin, in prep.). 
The IMF in each star cluster, $\xi_{M_\mathrm{ecl}}$, 
is normalised by the total star cluster
mass, 
\begin{equation}
M_\mathrm{ecl} =
\int_{m_\mathrm{low}}^{m_\mathrm{max}(M_\mathrm{ecl})}m\;\xi_{M_\mathrm{ecl}}(m)\;
\mathrm{d}m\;. 
\end{equation}
Finally, the local integrated galactic initial mass function (LIGIMF)
can be calculated by locally adding up all stars in all newly formed star clusters,
\begin{equation}
\xi_\mathrm{LIGIMF}(m,x,y)=
\int_{M_\mathrm{ecl,min}}^{M_\mathrm{ecl,max,loc}(x,y)}
\xi_\mathrm{M_\mathrm{ecl}}(m)\;\xi_\mathrm{LECMF}\;(M_\mathrm{ecl},x,y)\;
\mathrm{d}M_\mathrm{ecl}\;.
\end{equation} 
 
The surface density of ionising photons emitted by all new stars at the position
$x,y$  is 
\begin{equation}
  N_\mathrm{ion,\delta
    t}(x,y)=\int_{m_\mathrm{low}}^{m_\mathrm{max}}
\xi_\mathrm{LIGIMF}(m,x,y)\;N_\mathrm{ion,\delta
    t}(m)\;\mathrm{d}m\;,
\end{equation}
where $N_\mathrm{ion,\delta t}(m)$ is the total number of 
ionising photons\cite{pflamm-altenburg2007d}
emitted by a star with  mass
$m$ in time  $\delta t$. 
The H$\alpha$-luminosity  surface density then follows from 
\begin{equation}
\Sigma_\mathrm{H\alpha}(x,y)=3.02\times
  10^{-12}\;\mathrm{erg}\;N_\mathrm{ion,\delta t}(x,y)/\delta t\;.
\end{equation}

{\large\bfseries\noindent\sloppy 
  \textsf{ii) Photon leakage}}

In order to correct for the leakage of ionising radiation from H~{\sc
ii} regions we use a comparative study\cite{ferguson1996a} of the
H$\alpha$ radiation of well defined H~{\sc ii} regions and their
embedding diffuse ionised gas (DIG). This study of the two galaxies
NGC~247 and NGC~7793 shows hat the radial and azimuthal surface
luminosity of the DIG is highly correlated with bright H~{\sc ii}
regions and that the required power to sustain the DIG can only be met
by the ionising radiation from massive star formation.  In faint
H$\alpha$ luminosity regions the luminosity contribution of the DIG
can exceed the one by distinct H~{\sc ii} regions.  We use the
published values of the H$\alpha$ surface luminosities of H~{\sc ii}
regions only and H~{\sc ii} regions plus DIG in different regions of
these galaxies to construct a correction formula. Supplementary
Fig.~\ref{supp_fig_leakage} shows the fraction of the H$\alpha$
luminosity due to H~{\sc ii} regions of the total (H~{\sc ii} + DIG)
H$\alpha$ luminosity as a function of the H$\alpha$ luminosity in
H~{\sc ii} regions.  Two interpolating functions are used to describe
the resultant relation.  With
\begin{equation}
  y = \log_{10}\left(\frac{\Sigma_\mathrm{HII}}
    {\Sigma_\mathrm{HII~+~DIG}}\right)
\end{equation}
and
\begin{equation}
  x = \log_{10}\Sigma_\mathrm{HII}
\end{equation}
the high luminosity region of NGC~7793 and all data points of
NGC~247 can be easily interpolated (solid line) by
\begin{equation}
  \label{supp_eq_corr}
  y(x) = 0.22\;(x-32.3)\;.
\end{equation}
NGC~7793 alone can be interpolated (dashed line) by
\begin{equation}
  \label{supp_eq_corr2}
  y(x) = 0.22\;(x-32.3)-\exp(-(x-28.2))\;.
\end{equation}
The total surface luminosities are then given by
\begin{equation}
  \label{supp_eq_corr3}
  \log_{10}\left(\Sigma_\mathrm{HII~+~DIG}\right) = x - y(x)\;.
\end{equation}
Luminosities brighter than 10$^{32.3}$~erg~s$^{-1}$~pc$^{-2}$ need no
correction. The influence of photon leakage from H~{\sc ii} regions
for the Kennicutt-Schmidt diagram of Fig.~\ref{kennicutt_schmidt_fig}
is shown in Supplementary Fig.~\ref{supp_fig_ks}.  The corrected
values (small black squares) are obtained by applying Supplementary
equation~(\ref{supp_eq_corr3}) on the original values (open circles). 
The steeper
($N=1.4$) $\Sigma_\mathrm{H\alpha}$-$\Sigma_\mathrm{gas}$ relation
left of the H$\alpha$ cutoff and the position of the cutoff still
persist.  The lines are as described in the main text.  Correcting for
photon leakage therefore does not affect our results in any way.

{\large\bfseries\noindent\sloppy 
\textsf{iii) Local applicability of the IGIMF theory}}

Recently it has been demonstrated that the concept of clustered star
formation has a significant influence on the appearance of galaxies in
H$\alpha$\cite{pflamm-altenburg2007d}. The galaxy-wide H$\alpha$
luminosity scales non-linearly with the underlying total
star-formation rate (SFR). As a first consequence it has been shown
that a sample of the Sculptor group of dwarf irregular galaxies have
much higher SFRs than previously deduced and that their SFRs scales
approximately linearly with their gas mass. The sample consists only
of 11 galaxies and the validity of the obtained star formation
relation might be questionable. But a subsequent investigation of a
sample of 205 local star forming galaxies confirms this finding and
reveals a strictly linear relation between the SFR of a galaxy and its
total gas mass over five orders of magnitude in mass (Pflamm-Altenburg
and Kroupa, in preparation). Furthermore, the local UV-based star
formation law integrated over the disc of a galaxy is in perfect
agreement with global H$\alpha$ based and IGIMF corrected star
formation law.
Therefore, dwarf galaxies can be described very well with the
IGIMF-theory, ie. by the concept of clustered star formation. For a
radius of 3~kpc of a dwarf galaxy the reference area is about
28.3~kpc$^2$. In this context ''local'' regions in a disc galaxy are
annuli. E.g. an annulus between 10 and 11 kpc has an area of
66~kpc$^2$, much larger than a dwarf galaxy for which the IGIMF theory
is seemingly applicable. This means local regions in the context of
disc galaxies have test areas larger than dwarf galaxies which we have
mentioned to be excellently described by the IGIMF theory, i.e. by the
concept of clustered star formation.

\clearpage
{\large\bfseries\noindent\sloppy 
\textsf{iv) Sizes of outer disc star forming regions}}

The basis of the LIGIMF theory is that X-UV discs result from local
''stellar IMF biases''. The H$\alpha$-luminosity---star-cluster-mass
relation becomes non-linear\cite{pflamm-altenburg2007d} for cluster
masses $\le$~3000~$M_\odot$. The LIGIMF theory predicts that
inner-disc star-forming complexes sample the LECMF up to higher upper
star-cluster mass limits than outer-disc star-forming complexes. As UV
is a star formation tracer much less sensitive to the LIGIMF-effect
(i.e. the supression of OB-star numbers in low-SFR regions), 
the UV/H$\alpha$ luminosity ratio in the inner disc is smaller than in
the outer disc. Thus, a basic requirement of the LIGIMF theory is that
star forming regions in the inner disc are typically larger or more
massive than the outer disc ones.  A colour-magnitude diagram of UV
knots observed in M83 differentiates into outer and inner disc UV
knots (Fig. 3 in ref.\citeafterref{thilker2005a}).  UV knots younger
than 100~Myr have FUV-NUV colours
(Fig. 2 in ref.\citeafterref{zaritsky2007a}).
It can clearly be seen that the typical young ($\le20$~Myr) UV
outer-disc knots of M83 have masses $\le$ 1000~$M_\odot$ and the main
bulk lies at $\approx$ 500~$M_\odot$ while the inner knots
extend to much higher masses.  Only these star forming clusters
($\le$20~Myr old) are of interest as only they contribute to H$\alpha$
emission.  Only one outer-disc object stands out with
10$^4$~$M_\odot$. The presence of such a massive star cluster in the
outer region is not forbidden statistically (see next Supplementary
Section).  Young inner-disc objects populate the CMD densely up to
5000~$M_\odot$ and have masses which are typically significantly
larger than outer-disc UV knots. Outer-disc knots which are much
larger than 1000~$M_\odot$ are older than 20~Myr and can not be
considered in the context of the H$\alpha$ cutoff. But at these
somehwat later ages the inner-disc knots are also more
massive. Summarising: at any time the typical star forming regions are
less massive in the outer disc than in the inner disc. Additionally, the
median FUV luminosity of the inner-disc population is~3 times greater
than in the outer region\cite{thilker2005a}.  And outer-disc complexes
appear systematically fainter than their inner-disc
counterparts\cite{thilker2005a}.  Furthermore, the M83 FUV luminosity
function of outer-disc stellar complexes is steeper than for the
inner-disc population. A similar trend is reported\cite{lelievre2000a}
for the H$\alpha$ luminosity function of H~{\sc ii} regions in
NGC~628.  This is in full agreement with the concept of a varying
upper mass limit of the LECMF [equation~(\ref{eq_m_ecl_max_loc})].  If
a galaxy is divided into two parts, an inner and an outer part, then
the LECMF of the respective part has to be integrated over the whole
part of the galaxy to get the ECMF of each part separately.  The inner
part consists of regions where the LECMF has high local upper cluster
mass limits.  The outer part consists of regions where the LECMF has
only low or intermediate local upper cluster mass limits. 
Thus, solving the surface integral over the LECMF it follows
that the ECMF in the outer part is steeper than in the inner
part. This is the same integration effect as for galaxies with a low
SFR having steeper IGIMFs than galaxies with a high
SFR\cite{pflamm-altenburg2007d}.  The same holds for the
FUV-luminosity and H~{\sc ii}-region luminosity function as
observed. These observations are an indirect evidence that the upper
mass limit of the LECMF decreases towards galactic regions with lower
star formation surface densities.

\clearpage
Also the finding that 5\%--10\% of the UV knots in the extended disc
of NGC~628 have associated H$\alpha$ emission\cite{zaritsky2007a} is
in-line with the LIGIMF theory. This ratio of H$\alpha$ emitting
sources to non-emitting sources is entirely reasonable given the age
estimates of $\approx$20~Myr for H~{\sc ii} regions and
$\approx$400~Myr for GALEX knots, i.e. 20/400 = 5~\%.  As explained
above, an H$\alpha$-under-luminous star cluster is not the same as an
H$\alpha$-non-luminous cluster.  Each young UV source ($\le$20~Myr)
can appear as an H$\alpha$ source. The crucial point is that the UV
and H$\alpha$ luminosities of star clusters of mass
$\le$3000~$M_\odot$ scale differently with the star cluster
mass\cite{pflamm-altenburg2007d}.  Therefore, these
observations\cite{zaritsky2007a} are in agreement with the LIGIMf
theory.

{\large\bfseries\noindent\sloppy 
\textsf{v) Local maximum cluster mass}}

The presence, for example in outer-disc regions of galaxies, of
outstanding star cluster masses is not forbidden, but expected. To
demonstrate this we take the cluster mass probability distribution
density, $f(M)=k\;M^{-\beta}$, between the lower, $M_\mathrm{l}$, and
an upper limit, $M_\mathrm{u}$, normalised by
\begin{equation}
  1 = \int_{M_\mathrm{l}}^{M_\mathrm{u}} f(M)\;\mathrm{d}M\;.
\end{equation}
$N$ clusters are drawn from this distribution function with $\beta=2$,
$M_\mathrm{l}=5\;M_\odot$, and $M_\mathrm{u}=10^7\;M_\odot$, the mass
of the most massive star cluster of this set is stored.  This
experiment is repeated several times. We then get a distribution of
the most massive star cluster of a set of $N$ star clusters. In
general, the distribution function of the $i$-th massive star cluster
of a set of $N$ star clusters can be constructed as follows: the
probability that the i-th massive star cluster lies in the mass range
from $M$ to $M+\mathrm{d}M$ is $f(M)$. There are ${N\choose
1}=N$ possible realisations to choose the i-th massive star cluster
out of $N$.  The probability that $i-1$ more-massive star clusters lie
in the mass range from $M$ to $M_\mathrm{u}$ is
$\left(\int_{M}^{M_\mathrm{u}}f(M')\;\mathrm{d}M'\right)^{i-1}$.
There are
$N-1\choose i-1$ possible realisations to choose $i-1$ clusters out of
the remaining $N-1$ clusters. Finally, the probability that $N-i$
less-massive star clusters lie in the mass range from $M_\mathrm{l}$
to $M$ is
$\left(\int_{M_\mathrm{l}}^{M}f(M')\;\mathrm{d}M'\right)^{N-i}$. There
are ${N-i\choose N-i} = 1$ possible realisations to choose $N-i$
clusters out of the remaining $N-i$. Thus, the total distribution of
the i-th massive star cluster is given by
\begin{equation}
  p_{i,N}(M)=N {N-1\choose i-1}
  \left(\int_{M_\mathrm{l}}^{M}f(M)\;\mathrm{d}M\right)^{N-i}
  f(M)
  \left(\int_{M}^{M_\mathrm{u}}f(M)\;\mathrm{d}M\right)^{i-1}\;.
\end{equation}
For the most-massive star cluster ($i=1$) and a single part power law
for the mass function the distribution simplifies to
\begin{equation}
  \label{supp_eq_mmax_analytical}
  p_{1,N}(M)=N k 
  \left(\frac{k}{1-\beta}(M^{1-\beta}-M_\mathrm{l}^{1-\beta})\right)^{N-1}\;
  M^{-\beta}\;,\; \mathrm{for}\;\beta \not= 1.
\end{equation}
The distribution of the most-massive star cluster is plotted in
Supplementary Fig.~\ref{supp_fig_mmax} for $N=10,100,1000,10000$
(drawing 10$^6$ times from $f(M)$).  Shown are the distributions
obtained from Monte-Carlo simulations (grey histograms) and the
analytical distributions (black curves) calculated using Supplementary
equation~(\ref{supp_eq_mmax_analytical}).  If a test area in the inner
region of the galaxy with a high star formation rate is specified one
would expect, for example, 1000 clusters per galactic test area and
the expected most-massive star cluster and thus the upper mass limit
of the LECMF in a typical inner region would be about
10$^4$~$M_\odot$.  In a test area of equal size in the outer regions
were the star formation rate is lower by a factor of ten the expected
number of star clusters per test area is reduced by a factor of ten,
too.  The expected most-massive star cluster would have a mass of
slightly less than 1000~$M_\odot$, while a 10$^4$~$M_\odot$ star
cluster is unlikely but possible.
Thus, from the statistical
point of view, the typical star forming regions in the outer galaxies
have to be of lower mass than in the inner regions with a much higher
star formation rate.

{\large\bfseries\noindent\sloppy 
  \textsf{vi) Fitting formula}}

The relation between the theoretical H$\alpha$ surface luminosity and
the gas surface density 
can be very well fitted (Fig.~\ref{kennicutt_schmidt_fig}) 
by a polynomial of fifth order,
\begin{equation}\label{fit_fct_eq}y=a_0+a_1x+a_2x^2+a_3x^3+a_4x^4+a_5 x^5\;,\end{equation} 
where
\begin{equation}y=\log_{10}\left(\Sigma_\mathrm{gas}/M_\odot\mathrm{pc}^{-2}\right)\;\;\;,\;\;\;x=\log_{10}\left(\Sigma_{\mathrm{H}\alpha}/\mathrm{erg}\;\mathrm{s}^{-1}\;\mathrm{pc}^{-2}\right)\;.\end{equation}
The coefficients are listed in Supplementary Table~\ref{coeff_tab}. 
The underlying 
theoretical  star formation law [$N=1$ in Supplementary 
equation~(\ref{supp_eq_sfr_law})] is
\begin{equation}\label{eq_ligimf_sfr_law}\Sigma_\mathrm{SFR}=3.27\cdot 10^{-10}\;\mathrm{yr}^{-1}\;\Sigma_\mathrm{gas}\;.\end{equation}
By combining these two equations the LIGIMF theory allows a convenient
and accurate conversion of an observed H$\alpha$ 
luminosity surface density into a true star-formation-rate 
surface density. 

\clearpage
 
\def\tablename{{\bf\sffamily Supplementary Table}}

\setcounter{table}{0}

\begin{table}
  \begin{center}
    {\bf Fit-coefficients}\\
    
    \vspace{0.3cm}
    \begin{tabular}{|c|c|}
      \hline
      $a_0$  & -136.388\\
      \hline
      $a_1$  & 29.3544\\
      \hline
      $a_2$  & -2.52783\\
      \hline
      $a_3$  & 0.108099\\
      \hline
      $a_4$  & -2.29326e-03\\
      \hline
      $a_5$  & 1.93394e-05\\
      \hline
    \end{tabular}\\
    \caption{Coefficients for the
      $\Sigma_\mathrm{H\alpha}$-$\Sigma_\mathrm{gas}$ fit.}{\label{coeff_tab}Coefficients of the fitting function  
      [Supplementary equation~(\ref{fit_fct_eq})] of the relation 
      between the H$\alpha$ surface
      luminosity and gas surface density
      (Fig.~\ref{kennicutt_schmidt_fig}).}
    \end{center}
\end{table}

\clearpage

\def\figurename{{\bf\sffamily Supplementary Figure}}

\setcounter{figure}{0}

\begin{figure}
  \begin{center}
    \includegraphics[width=0.7\textwidth]{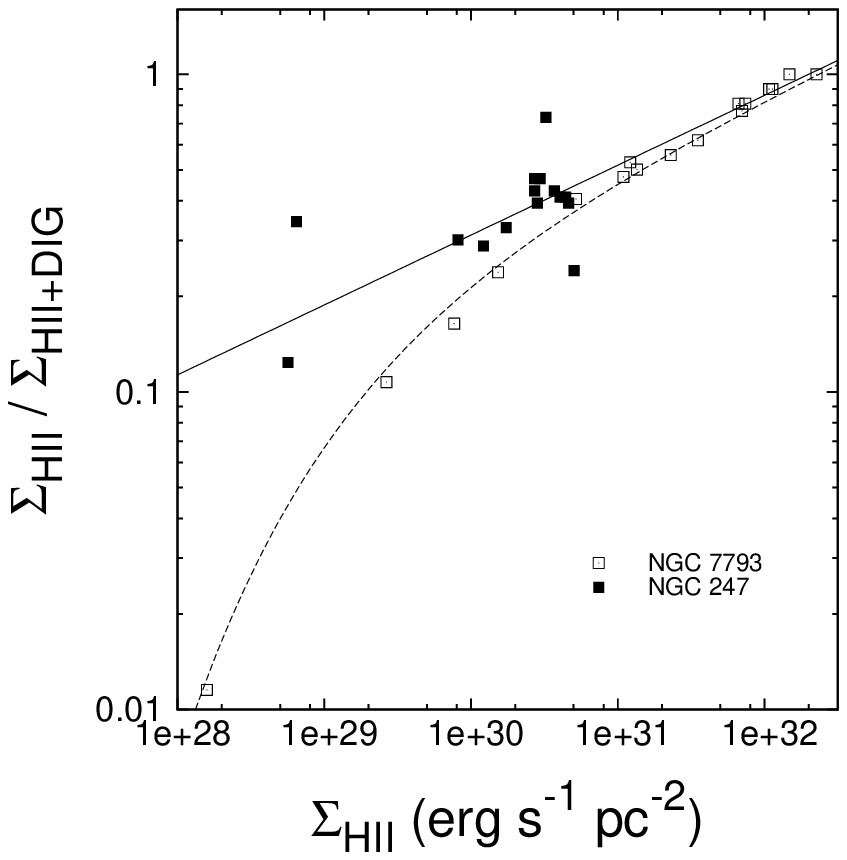}
  \end{center}
  \caption{Estimating the photon leakage from H{\sc ii} regions.}{\label{supp_fig_leakage}
    Ratio of the H$\alpha$-surface luminosity of H~{\sc ii} regions only
    and the total H$\alpha$-surface luminosity (H~{\sc ii} regions plus DIG) 
    versus 
    H$\alpha$-surface luminosity of H~{\sc ii} regions only, based
    on a comparative study\cite{ferguson1996a} 
    of the two disc galaxies NGC~7793 and NGC~247 in annuli at different
    galactocentric radii.
  }
\end{figure}

\clearpage

\begin{figure}
  \begin{center}
    \includegraphics[width=0.9\textwidth]{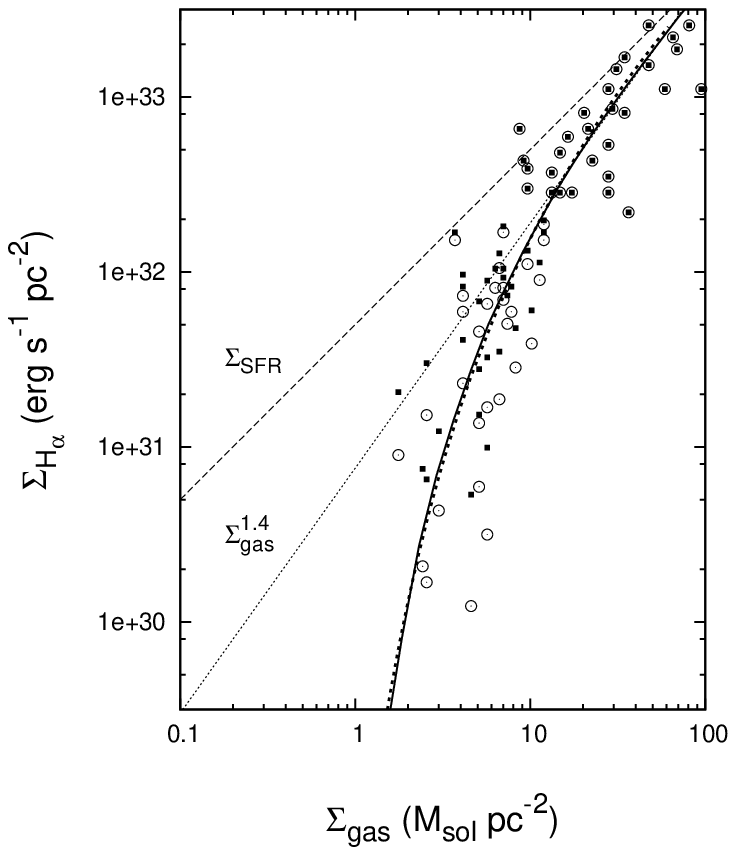}
  \end{center}
  \caption{Influence of photon leakage on the derived H$\alpha$ surface density.}{\label{supp_fig_ks}
    H$\alpha$-surface luminosity density corrected for photon leakage
    (small black squares) and uncorrected (open circles) versus 
    gas surface density of the same galactic annuli.
  }
\end{figure}

\clearpage

\begin{figure}
  \begin{center}
    \includegraphics[width=0.9\textwidth]{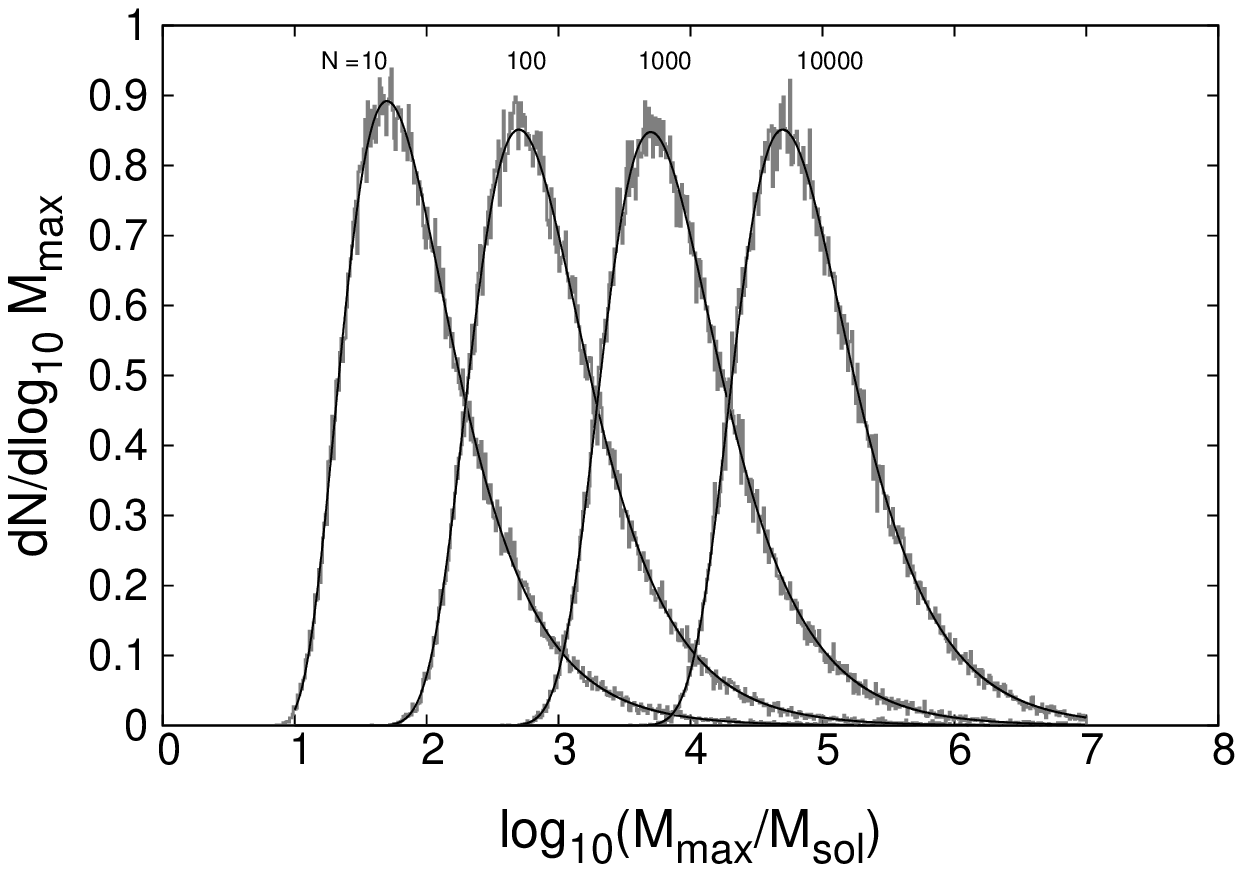}%
    \caption{Most-massive star cluster distribution.}{\label{supp_fig_mmax} Distribution of the most massive star
    cluster of a set of $N$ clusters from Monte-Carlo simulations 
    (grey histograms) and the analytical treatment (black lines).}
  \end{center}
\end{figure}

\end{document}